\input{aipcheck}

\documentclass[final,sort&compress]{aipproc}
\usepackage{amsfonts}
\usepackage{amsmath}
\usepackage{amssymb}
\usepackage{graphicx}

\layoutstyle{6x9}

\begin{document}

\title{Self-Organized Criticality  and  earthquakes    }


\author{Filippo Caruso}{
  address={NEST-CNR INFM - Scuola Normale Superiore, Piazza dei Cavalieri 7, I-56126  Pisa, Italy}
}

\author{Alessandro Pluchino}
{
  address={Dipartimento di Fisica e Astronomia and INFN, Via S. Sofia 64, 95123 Catania, Italy}
}

\author{Vito Latora}
{
  address={Dipartimento di Fisica e Astronomia and INFN, Via S. Sofia 64, 95127 Catania, Italy}
}
\author{Andrea Rapisarda}
{
  address={Dipartimento di Fisica e Astronomia and INFN, Via S. Sofia 64, 95127 Catania, Italy}
}
\author{Sergio Vinciguerra}{
  address={Dept. of Seismology and Tectonophysics,
INGV, I-00143 Roma, Italy}
}

\begin{abstract}
 We  discuss  recent results on  a new  analysis regarding 
models showing Self-Organized Criticality (SOC), and in particular on the 
OFC one. We  show that Probability Density Functions 
 (PDFs) for the avalanche size differences at different times have 
 fat tails with a q-Gaussian shape.
This behavior does not
depend on the time interval adopted and it is also found
when considering energy differences  between real earthquakes.
\end{abstract}

\maketitle

\section{Introduction}

In the study of earthquake dynamics,
the   Self-Organized Criticality (SOC)
paradigm proposed by Bak and coworkers \cite{Bak:1989} has been lengthly 
debated  during the last decade  in order to clarify the controversial   earthquakes   predictability 
\cite{debate}. In this short paper, we discuss recent results \cite{Filippo:2006,Filippo:2007} 
and we show that it is possible  to reproduce statistical features of
earthquake catalogs within a SOC scenario, if one takes 
into account long-range interactions. Here we consider 
 the  dissipative Olami-Feder-Christensen model \cite{Olami:1992} on a \textit{small world}
topology  and we  show that the
Probability Density Functions (PDFs) for the avalanche size
differences at different times have fat tails with a q-Gaussian
shape \cite{Tsallis:2005} only if finite-size
scaling (FSS)  is present. This behavior does not depend on the time
interval adopted and it is  found also after reshuffling the data.
Similar results have been obtained   
if energy differences between real earthquakes
are considered.

\begin{figure} [ht]
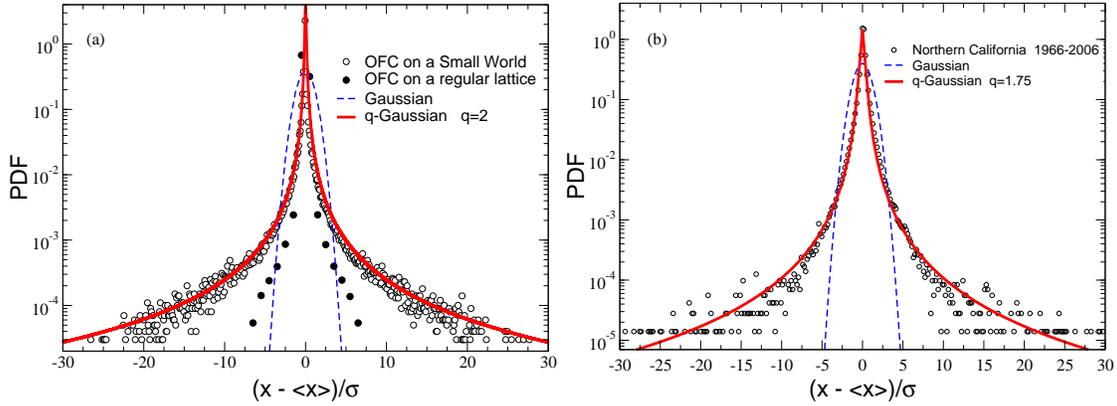

\includegraphics[width=7.3cm]{f1.eps}
\includegraphics[width=7.3cm]{f2.eps}
\caption{\label{expM} 
(a) PDF of the avalanche size differences (returns)
$x(t)=S(t+1)-S(t)$ for the OFC model on a small world topology
(critical state, open circles) and on a regular lattice (non
critical state, full circles). The  curves
are normalized to the
standard deviation $\sigma$.  The first curve can be well fitted by 
a q-Gaussian (full line) with an exponent $q\sim2.0\pm0.1$ \cite{Tsallis:2005}. A
standard Gaussian (dashed line) is also reported for comparison.
All the  curves were normalized so to have unitary area. 
 (b) PDFs of the energy differences $x(t)=S(t+1)-S(t)$ are
shown  for the
Northern California Catalog\cite{nc}. Data can be well reproduced  by a q-Gaussian fit 
(full line) with $q\sim1.75\pm0.15$.
A standard Gaussian is also plotted as dashed line. }
\end{figure}

\section{Model and results}

The Olami-Feder-Christensen (OFC) model \cite{Olami:1992}   is one of the
most interesting models displaying Self-Organized Criticality.
Despite of its simplicity, it exhibits a rich phenomenology
resembling real seismicity. In its original version the OFC model
consists of a two-dimensional square lattice of $N=L^2$ sites,
each one connected to its $4$ nearest neighbours and carrying a
seismogenic force represented by a real variable $F_i$, which
initially takes a random value in the interval $(0,F_{th})$. In
order to mimic a uniform tectonic loading all the forces are
increased simultaneously and uniformly, until one of them reaches
the threshold value $F_{th}$ and becomes unstable $(F_i \geq
F_{th})$. The driving is then stopped and an ''earthquake'' (or
avalanche) starts, i.e. each unstable  site $i$ releases a part of its  force,
proportional to $F_i$ to its four neighbours. 
The number of topplings during an avalanche defines its size $S$,
while the dissipation level of the dynamics is controlled by the parameter
$\alpha \in[0,0.25]$. The model is conservative
if $\alpha=0.25$, while it is dissipative for $\alpha<0.25$.
In the following  we consider the
dissipative version of the OFC model with $\alpha=0.21$ 
on a regular lattice with $L=64$ and open boundary conditions
(i.e. we impose $F=0$ on the boundary sites). In order to
improve the model in a more realistic way, we introduced a small
fraction of long-range links in the lattice so to obtain a \textit{small
world }topology. Just  a few long-range edges create
short-cuts that connect sites which  otherwise would be much
further apart. Long-range connections  allow the system to
synchronize and to show both FSS and universal
exponents as was  shown in Ref.\cite{Filippo:2006}.
Furthermore, a small world topology
is expected to model  more accurately earthquakes spatial correlations,
taking into account long-range as well as short-range seismic
effects.  In our version of the OFC
model the links of the lattice are rewired at random with a
probability $p$. The 
transition to obtain small world features and criticality is
observed at  $p=0.02$ \cite{Filippo:2006}.
In Ref. \cite{Filippo:2007} we calculated  the distribution of the avalanche size
time-series $S(t)$ for the OFC model on a small world topology 
and on a regular lattice. In our case the time
 $t$ is  a progressive discrete index
labelling successive events and is analogous to the \textit{   "natural time" } successfully used
in Ref. \cite{Varotsos:2005}.
In order to  have  a good statistics, we considered up to $10^9$  avalanches in our numerical experiments.
In both cases, as shown in Fig.1(a) of  Ref. \cite{Filippo:2007}, the  PDFs  follow  a power-law decay
$y\sim S^{-\tau}$ with a slope $\tau=1.8\pm0.1$ even if real criticality is present only for the small world topology \cite{Filippo:2006}. In the last years
SOC models  have been  intensively studied considering  time
intervals between avalanches in the  critical regime. Here we 
discuss  a different approach which  reveals
interesting information on the eventual criticality  of the model
under examination. 
We focus our
attention  on the \emph{''returns''} $x(t)=S(t+\Delta)-S(t)$, i.e. on
the differences between avalanche sizes calculated at time
$t+\Delta$ and at time $t$, $\Delta$ being a  discrete time
interval.
The resulting signal is extremely intermittent at
criticality, since successive events can have very different
sizes.
On the other hand, if  the system is not in a critical
state, this intermittency character is very reduced. 

In Fig.1 (a)
we plot as open circles the Probability Density Function (PDF) of
the returns $x(t)$ (with $\Delta=1$) obtained for the critical  OFC
model on small world topology. The returns are normalized in order
to have zero mean and unitary variance. The curves  reported have
also unitary area. A behavior very different from a Gaussian
shape (plotted as dashed curve) is observed: the PDF is  very peaked
and exhibits fat tails. On the other hand, for the model on regular lattice, even if
 power laws for  the avalanche  size are found,  the model  is
not critical  since no FSS  is observed
\cite{Filippo:2006}.  In this latter case  no fat tails emerge, although a sensible
departure from Gaussian behavior is evident (see full
circles). These findings thus indicate a new powerful way for characterizing
the presence of criticality.
Another remarkable  feature which we have  found is that such a behavior does not
depend on the interval $\Delta$ considered for the  avalanche size
difference.  Even after  reshuffling the data, i.e. changing in a random way the
time order of the avalanches,  no change in the
PDFs was  observed.  Moreover, the data reported  in Fig.1 (a) for the critical OFC model
on a small world can be well fitted by  a \textit{q-Gaussian }
curve  $~~f(x)= A[1-(1-q){x^2}/B]^{1/(1-q)}~~$ which is typical of Tsallis
$q$-statistics \cite{Tsallis:2005}. This function generalizes the standard
Gaussian curve, depending on the parameters $A,B$ and on the
exponent $q$.  For  $q=1$ the normal distribution is obtained
again, so $q \ne 1$ indicates a departure from Gaussian
statistics. The q-Gaussian curve, reported as full line,
reproduces very well the model behavior in the critical regime,
yielding in our case  a  value of $q=2.0\pm0.1$.

In order to compare these theoretical results with real earthquakes data,
we repeated the previous analysis for several catalogs \cite{Filippo:2007}.
Here   we discuss  the
Northern California catalog  for the period 1966-2006 \cite{nc}. The latter is
a very extensive and complete
seismic data set on one of the most active and studied faults on the Earth, i.e. the San
Andreas Fault. The
 total number of earthquakes considered is almost 400000.
Actually 
the energy, and not the magnitude, is the quantity  which should
be considered equivalent to the avalanche size in the OFC model.
Therefore we  studied the  quantity $S=exp(M)$, where $M$ is  the
magnitude of a real earthquake. This quantity is simply related to
the energy dissipated in an earthquake, the latter being an increasing
exponential function of the magnitude.

In Fig.1 (b)  we consider the released energy $S$ for the Northern California catalog
and we plot the 
PDFs of the corresponding  returns $x(t)=S(t+\Delta)-S(t)$ (with
$\Delta=1$).
Also for  real data \textit{t } is a progressive discrete index labelling
 successive events.
As  for the
critical OFC model previously discussed, fat tails and
non-Gaussian probability density functions are observed. In both
cases the experimental points can be fitted by a q-Gaussian curve,
obtaining an exponent $q=1.75\pm0.15$, a value which is
compatible, within the errors, to that one found for the OFC
model.
In Ref. \cite{Filippo:2007}
also the   world  catalog was considered  with similar results. 
As for the OFC model also 
for the real earthquakes data, by varying  the
interval $\Delta$ of the energy returns $x$, or by reshuffling the
time-series $S(t)$, no change in the  PDFs was  observed.
The   above results give further  support to the argument  that a  SOC mechanism underlies 
earthquake  dynamics and that, actually, although the  system is in a  strongly correlated regime in 
space, there is no correlation in time between 
the magnitude of successive events, as also explained  by means of a simple
analytical derivation in  Ref.  \cite{Filippo:2007}.

\section{Conclusions}

We have discussed  a new type of analysis which is able to discriminate in a quantitative way real SOC
dynamics. The method, here applied  to the OFC  model and to real  earthquakes data, gives further  support 
to the argument  that seismicity can be explained within a dissipative self-organized criticality scenario when 
long-range interactions are taken into consideration.  


\begin{theacknowledgments}
We acknowledge
financial support from the PRIN05-MIUR project \textit{Dynamics and thermodynamics of systems with long-range 
interactions.}
\end{theacknowledgments}



\begin{thebibliography}{9}

%



\bibitem{Bak:1989}
P. Bak,  C. Tang,  \emph{J. Geophys. Res. }{\bf 94}, 15635--15637 (1989). 

\bibitem{debate}
Nature debates, {\it Is the reliable prediction of individual
earthquakes a realistic scientific goal?} (1999), see 
\textit{http://www.nature.com/nature/debates/earthquake}


\bibitem{Filippo:2006}
F.~Caruso, V.~Latora, A.~Pluchino, A.~Rapisarda, B.~Tadic,
\emph{Eur. Phys. Journ. B }  \textbf{50}, 243-247 (2006).

\bibitem{Filippo:2007}
F.~Caruso, A.~Pluchino, V.~Latora, S.~Vinciguerra, A.~Rapisarda,
\emph{Phys. Rev. E }  \textbf{75}, 055101(R)  (2007).

\bibitem{Olami:1992}
Z.~Olami, H.J.S.~Feder,  K.~Christensen, \emph{ Phys. Rev. Lett. } {\bf
68}, 1244--1247 (1992).

\bibitem{Tsallis:2005}
C.~Tsallis, M~Gell-Mann,   Y.~Sato, Europhys. News
\textbf{36}, 186--189 (2005) and refs. therein.

\bibitem{Varotsos:2005}
P. A.  Varotsos, N.V. Sarlis, H.K. Tanaka and E.S. Skordas,\emph{ Phys. Rev. E} {\bf  72},
041103 (2005) and refs.  therein.

\bibitem{nc}
The  data of the Northern California earthquakes were taken from \textit{http://www.ncedc.org/ncedc/catalog-search.html}

\end{thebibliography}
\end{document}